\documentclass[12pt]{article}
\usepackage{natbib}
\usepackage[]{hyperref}  
\usepackage{bbm, bm,float}
\usepackage{graphicx}
\usepackage{algorithm}
\usepackage[noend]{algpseudocode}
\usepackage[figuresright]{rotating}
\usepackage{authblk}
\usepackage{booktabs}
\usepackage{placeins}

\usepackage{amsfonts}       
\RequirePackage{amsthm,amsmath,amsfonts,amssymb}

\title{Ridge-penalized adaptive Mantel test and its application in imaging genetics}

\author[1]{Dustin Pluta}
\author[1]{Tong Shen}
\author[2]{Chuansheng Chen}
\author[3]{Gui Xue}
\author[1, 4]{Hernando Ombao}
\author[1,*]{Zhaoxia Yu}
\affil[1]{Department of Statistics, University of California, Irvine, USA}
\affil[2]{Department of Psychology and Social Behavior, University of California, Irvine, USA}
\affil[3]{State Key Laboratory of Cognitive Neuroscience and Learning and IDG/McGovern Institute for Brain Research, Center for Collaboration and Innovation in Brain and Learning Sciences, Beijing Normal University, Beijing, PR China}
\affil[4]{Biostatistics Group, King Abdullah University of Science and Technology, Saudi}
\affil[*]{Corresponding author: Zhaoxia Yu, zhaoxia@ics.uci.edu}

\date{\today}

\begin{document}
\maketitle 


\abstract[Summary]{We propose a ridge-penalized adaptive Mantel test (AdaMant) for evaluating the association of two high-dimensional sets of features. By introducing a ridge penalty, AdaMant tests the association across many metrics simultaneously. We demonstrate how ridge penalization bridges Euclidean and Mahalanobis distances and their corresponding linear models from the perspective of association measurement and testing. This result is not only theoretically interesting but also has important implications in penalized hypothesis testing, especially in high dimensional settings such as imaging genetics. Applying the proposed method to an imaging genetic study of visual working memory in health adults, we identified interesting associations of brain connectivity (measured by EEG coherence) with selected genetic features.}


\maketitle


\section{Introduction}
\label{sec1}
The primary goal of imaging genetics is to identify genes that might influence variation in neurophysiological function and neuroanatomical structure. Thus, imaging genetic studies typically collect neuroimaging and genetic data with hundreds of thousands of features per modality, while sample sizes are usually on the order of several dozen to a few thousand. In this scenario, both explanatory and response variables of interest are high-dimensional, such as thousands of gene expressions, millions of DNA variants, dozens to hundreds of EEG channels, and hundreds to millions of blood oxygen level dependent (BOLD) signals. Given how little is known about the relationship of neurological phenotypes with genetics, it is of interest to determine whether particular sets of features are significantly associated in aggregate. However, developing association tests that are sufficiently powered remains an important statistical challenge for the analysis of imaging genetics data. 

Mantel's test \cite{mantel1967} is one of the earliest formulations of a (ostensibly non-parametric) distance-based test for association of features from two observational modalities.  In the Mantel tes (MT), one computes a distance matrix between pairs of subjects in each modality, and tests for association of the two distance matrices. This simple approach is particularly appealing for two sets of high dimensional variables as it is flexible to choose any distance metric. Improvements have been made by the use of a Procrustean superimposition approach \cite{Peres-Neto2001Oecologia}, which is closely related to the double-centering transformation \cite{gower1966some} used in several articles, such as Wessel and Schork\cite{wessel2006generalized}, Salem et al.\cite{salem2010curve}, and Schork and Zapala\cite{schork2012statistical}. As described in Section 2, in some special scenarios, after double centering, distances are converted to similarities. Closely related tests have been developed from Escoufier's RV coefficient and its application to neuroimaging \cite{robert1976unifying, shinkareva2006classification}, the distance covariance (dCov) \cite{szekely2007measuring}, and the sum of powered score tests (aSPC) \cite{xu2017adaptive}. A detailed discussion of the connections of MT, the RV coefficient, and dCov is given in Omelka and Hudecová \cite{omelka2013comparison}. Similar methods have frequently been used for genetic association analysis, e.g.  Haseman and Elston\cite{haseman1972investigation}, Schaid et al.\cite{schaid2005nonparametric}, Beckmann et al. \cite{beckmann2005haplotype}, Wessel and Schork\cite{wessel2006generalized}, Salem et al. \cite{salem2010curve}, Schaid \cite{schaid2010genomic},  Mukhopadhyay et al.\cite{mukhopadhyay2010association}, Wei et al.\cite{wei2008u}, Goeman et al. \cite{ goeman2004global},  Schork and Zapala,\cite{schork2012statistical}. An excellent review of RV coefficient, its similar methods and extensions can be found in Josse and Holme \cite{josse2016measuring}.

In modern statistics, shrinkage methods play critical roles in achieving various statistical inference goals. In particular, in high dimensional settings, shrinkage methods are necessary for numerical stability, interpretation, preventing overfitting, and constructing powerful test. Ridge regression, which performs $L^2$ regularization, was originally proposed to deal with ill-conditioned or close to ill-conditioned problems \cite{tikhonov1943stability, horel1962applications, hoerl1970ridge}, and has since proven extremely useful in prediction problems in various scientific areas, such as predicting human complex traits and agricultural and breeding outcomes \cite{cule2011significance, hayes2001prediction, de2013prediction}. Ridge penalty is also widely used in neural network to reduce overfitting. This suggests that ridge penalization may yield useful association tests, but despite the existing rich literature on applications of $L^2$ penalized test e.g., Liu et al.\cite{liu2007semiparametric}, Lin et al. (2013)\cite{lin2013test}, Lin et al. (2016)\cite{lin2016test}, Cule et al.\cite{cule2011significance}, the role that ridge penalty plays in hypothesis testing has been rarely studied. For prediction in deep learning and estimation, the results are mixed. For example, the optimal ridge penalty is found to be positive when the coefficients are generated from a distribution \cite{hastie2019surprises} whereas the optimal ridge penalty might be negative when the coefficients are fixed \cite{kobak2018optimal}. However, in practice the true model is never known. Motivated by these challenges, in this article, we propose the ridge-penalized adaptive Mantel test (AdaMant) for evaluating the association of two high-dimensional sets of features. By introducing a ridge penalty, AdaMant tests the association across many metrics simultaneously.  

The structure of  this article is as follows.  Section \ref{mantel} introduces a group of Mantel tests and describes how ridge penalization, which induces a ridge kernel and distance, connects Euclidean distances to Mahalanobis distances and their corresponding linear models. In Section \ref{adamant}, we propose the adaptive Mantel test and describe its implementation. Section \ref{simulations} evaluates the performance of the test using simulations. Section \ref{application} presents results showing significant associations between brain connectivity (measured by EEG coherence) and genetic features from a working memory study of 350 college students. Section \ref{discussion} concludes with a discussion of generalizations and directions for future work.  

\section{Mantel Test and Linear Models}\label{mantel}
The framework of Mantel test is extremely general, and encompasses many well-known association tests, such as the RV coefficient \citealp{robert1976unifying}, the distance covariance \citealp{szekely2007measuring}, and the sum of powered score tests \citealp{xu2017adaptive}. Consider data $(X_i, Y_i) \in \mathbb{R}^p \times \mathbb{R}^q, i = 1, \dots, n$, where $n$ is the number of observations (participants), $p$ is the number of independent variables such as $500K$ SNPs, $q$ is the number of dependent variables, such as 190 pairwise theta band coherence from 20 EEG channels. The corresponding data matrices $X_{n\times p}$ and $Y_{n\times q}$ have been column-centered. To state the general form of the Mantel test, let $\mathcal{K}^{\textbf{X}}(\cdot, \cdot)$ be a positive semi-definite (p.s.d) kernel function on $\textbf{X} \times \textbf{X}$, with corresponding Gram matrix $H$ defined by $H_{ij} = \mathcal{K}^{\textbf{X}}(X_i, X_j)$, and define $\mathcal{K}^{\textbf{Y}}$ and $K$ similarly using $Y$.  The Mantel test statistic for these kernels is equivalent to $\text{tr}\left(HK\right)$. The reference distribution under the null hypothesis of no association between similarities measured by $\mathcal{K}^{\textbf{X}}$ and $\mathcal{K}^{\textbf{Y}}$, can be obtained from the observed features $X$ and $Y$ by permuting the observation labels for one set of features and calculating the empirical reference distribution. Equivalently, one can hold one matrix fixed, say \(H\), and simultaneously permute the rows and columns of \(K\).

Let $X_i$ and $X_j$ be the vector of independent variables for subjects $i$ and $j$, respectively. One simple kernel (similarity) metric is their inner product $\langle X_i,X_j\rangle=X_i^TX_j$, which is commonly known as the linear kernel in kernel regression or the dot product similarity. It is natural to generalize it by applying a p.s.d weight matrix $\mathcal W$: $\mathcal{K}_{\mathcal{W}}(X_i, X_j) = \langle X_i, X_j\rangle_{\mathcal W} = X_i^T \mathcal{W} X_j$.  Particular choices of $\mathcal{W}$ yield Mantel-type tests that are corresponding to different distance measures. Choosing $\mathcal W = I_p$ (the $p \times p$ identity matrix) gives the standard Euclidean inner product, with Gram matrix $ XX^T$, i.e., the linear kernel. Setting $\mathcal{W} = (X^TX)^{-}$, gives Gram matrix $X(X^TX)^{-}X^T,$  which is recognizable as the projection matrix into the column space of $X$. Pre-conditioning the Mahalanobis weight matrix as $\mathcal{W} =  (X^TX + \lambda I_p)^{-1}$ gives Gram matrix $X(X^TX + \lambda I_p)^{-1}X^T.$ As shown in the Appendix A.1, the above similarity matrices are identical to their double-centered squared distances. For example, let $C=\mathbf{I}_n-\mathbf{1}\mathbf{1}^T/n$ (the centering matrix) and let $D^2_\lambda$ be the squared distance matrix whose $(i,j)$th element is the squared distance between subjects $i$ and $j$: $$D^2_{\lambda,(i,j)}=(X_i-X_j)^T (X^TX+\lambda \mathbf{I}_p)^{-1}(X_i-X_j).$$ We have $X (X^TX+\lambda \mathbf{I}_p)^{-1} X^T = -\frac{1}{2} \mathcal{C} D^2_\lambda \mathcal{C} $. Thus, it is sensible to use Euclidean and Mahalanobis similarity, respectively, to refer to $XX^T$ and $X(X^TX)^{-}X^T$.

\subsection{Score Statistics in Multivariate Linear Models}
We next show that Mantel tests using the weighted matrices introduced above are closely related to score statistics in linear models.  Rao's score test \cite{rao1948large} is based on the gradient of the log-likelihood evaluated when the null hypothesis is true. It is locally most powerful and asymptotically equivalent to the Wald and likelihood ratio tests. It is particularly attractive for high-dimensional inference since it does not require the estimation of the parameters to be tested, greatly reducing the computation required compared to alternative approaches. This also makes it convenient to calculate the null distribution for the score test statistic via a permutation procedure, which forms the basis of the adaptive Mantel test algorithm proposed in Section \ref{algorithm}.

\subsubsection{Adjust for Covariates}
To simplify the presentation of mathematical derivations, we assume that $X$ only consists of variables of direct interest and covariates such as age and gender have been adjusted for. In many challenging problems, such as imaging genetics, both the outcome and explanatory variables are high-dimensional but the number of covariates to be adjusted for is usually small. These covariates are not of primary interests. One simple method to handle covariates is to first compute the residuals by regressing the outcomes on the covariates and then replace the outcomes with the residuals. This type of of covariate adjustment in mixed-effects models has been commonly used \citealp{goeman2011testing, liu2007semiparametric}. A more accurate method is to use the restricted maximum likelihood estimates (REML)-type transformation (e.g., Ge et al.\citealp{ge2016multidimensional}). Through this article, we assume that the $n\times q$ outcome matrix $Y$ and the $n\times p$ matrix of explanatory variables $X$ have been covariate-adjusted. Note that $n$ is smaller than the number of observations if the REML-type of transformation method is use. 

\subsubsection{Linear Fixed and Random Effects Models}
The conventional linear model is a fixed effects model where the coefficients are treated as fixed but unknown parameters. When the number of explanatory variables $p$ is large relative to the sample size $n$, the design matrix is ill-conditioned or close to ill-conditioned and regularization methods are often applied.  An alternative is the random effects model which assumes that the coefficients are random variables from a normal distribution. The random effects model has been widely used to handle correlated data such as repeated measurements in longitudinal studies \cite{laird1982random}. Because it can naturally analyze a large number explanatory variables and account for correlation across observations, the variance components form of the random effects model has been used for many years in genetic studies to estimate the heritability of phenotypes from a sample of related or unrelated subjects. Consider the following two linear models.
\begin{enumerate}
	\item 
	\underline{Fixed Effects Model}:         $Y=X\beta+\epsilon$, where $\epsilon \sim N(0, \mathbf{I}_n, \Sigma),$ which can be equivalently expressed using vectorized $Y$: $vec(Y)\sim N(vec(X\beta), \Sigma \otimes \mathbf{I}_n)$, where $vec(\cdot)$ denotes the vectorization operator.  
	
	\item \underline{Random Effects Model}: $Y=X b+\epsilon$, where $\epsilon \sim N(0, \mathbf{I}_n, \Sigma)$	and  $b\sim N(0, \mathbf{I}_p, \Sigma_b)$. Equivalently, $vec(Y)\sim N(0,\Sigma_b \otimes XX^T + \Sigma \otimes \mathbf{I}_n)$
\end{enumerate}

Their null hypotheses are $\beta=\mathbf{0}$ and $\Sigma_b=0$, respectively. For the random effects model, the score equals a quadratic form minus its expectation, where the expectation is evaluated under the null hypothesis. Because the expectation equals the product of a function of $X$ and a function of $Y$, it is not informative for association testing. As a result, score tests are often conducted based on the quadratic form, e.g., Liu et al.\citealp{liu2007semiparametric}, Huang and Lin\cite{huang2013gene}, Goeman et al. 2004\cite{goeman2004global}, Goeman et al. 2011\cite{goeman2011testing}. Using the same strategy, we have the following result:\\

\noindent \textbf{Proposition 1}. Let $H_0=X(X^TX)^{-}X^T$, $K_0=Y(Y^TY)^{-}Y^T$, $H_\infty=XX^T$, and $K_\infty=YY^T$. The score statistic from the fixed effects model is proportional to $tr(H_0K_0)$.  When $\Sigma = \sigma^2 \mathbf{I}_q$ and $\Sigma_b=\sigma_b^2 \mathbf{I}_q$, which is always true when $q=1$, the score statistic from the random effects model is proportional to $tr(H_\infty K_\infty)$.  Asymptotically, their null distributions are chi-squared and mixture of chi-squared, respectively.

The proof is in Appendix A.2. \textbf{Remarks}. (1) When $\Sigma_b$ in the random effects model is unstructured, we have a score matrix because there is a score for each element of $\Sigma_b$. If we use its trace, i.e., the sum of its diagonal elements, the corresponding test statistic is still $tr(H_\infty K_\infty)$. The assumption of $\Sigma_b=\sigma_b^2 \mathbf{I}$ implicitly assumes that the coefficients in $b$ are exchangeable, i.e., the density of all permutations of the coefficients are the same. (2) If $\Sigma$ in the random effects model is unstructured, the score statistic involves $Y\Sigma^{-2}Y^T$, which is proportional to $Y(Y^TY)^{-2}Y^T$ when $\Sigma$ is replaced by a consistent estimator. This result is consistent with those presented in earlier work such as Maity et al.\citealp{maity2012multivariate}, Huang and Lin\cite{huang2013gene}. Although $\Sigma=\sigma^2\mathbf{I}$ is an oversimplification, when the dimension $q$ is large, it is costly to estimate $\Sigma$. The ridge penalized similarity introduced next might be considered as a compromise.

\subsubsection{Ridge Penalization and a Data-Augmentation Approach}
The ridge regression is a $L_2$-penalized form of the fixed effects model with the following penalized log-likelihood: 
\begin{equation}
    l_{\lambda_X}=-\frac{n}{2}log|\Sigma|- \frac{1}{2} tr[(Y-X\beta)\Sigma^{-1}(Y-X\beta)^T]-\frac{\lambda_X}{2} tr(\beta \beta^T),
    \label{eqn:pen_loglik}
\end{equation}

\noindent where $\lambda_X\ge0$ is the ridge parameter.

The ridge estimator of $\beta$ is  $\hat\beta_{\lambda_X} = {\arg \min}_\beta \left\{tr[(Y-X\beta)\Sigma^{-1}(Y-X\beta)^T]+\lambda_X tr(\beta \beta^T)\right\}$. Recall that ridge regression can be formulated by augmenting both $X$ and $Y$ (``12.3.1 Physical interpretations of ridge regression" in Sen and Srivastava\citealp{sen2012regression}). For example, the log-likelihood and ordinary least squares estimate (OLSE) based on the following augmented data is mathematically identical to the the log-likelihood in (\ref{eqn:pen_loglik}) and the ridge-penalized estimate $\hat\beta_{\lambda_X}$, respectively: $\tilde Y=[Y^T \vdots ~ \mathbf{0}_{q \times p}]^T, \tilde X = [X^T \vdots ~ \sqrt{\lambda} \mathbf{I}_p]^T$. Because ridge regression does not involve $\lambda_Y$, it is sensible to consider a symmetric treatment of $X$ and $Y$ when both of them are high-dimensional. Similar data augmentation strategies have been proposed in Hodges\citealp{hodges1998some} for the purpose of model diagnostics in hierarchical models. Consider two tuning parameters $\lambda_X>0, \lambda_Y > 0$. We propose to use the following augmented data

\begin{equation}
\tilde{Y}=\begin{pmatrix}Y\\ \mathbf{0}\\ \sqrt{\lambda_Y} I_q\end{pmatrix}, \tilde{X}=\begin{pmatrix}X\\ \sqrt{\lambda_X} I_p \\ \mathbf{0}\end{pmatrix}
\end{equation}

If we model $\tilde X$ and $\tilde Y$ with the same distributional assumptions as used to the fixed-effects model, we can define an augmented-data likelihood. Although it is not a real likelihood, nevertheless, one can still define score statistics. By \textbf{Proposition 1}, the corresponding score statistic for $\beta=0$ is proportional to $tr(\tilde H \tilde K)$, where $\tilde H=\tilde X (\tilde X^T \tilde X)^{-1} \tilde X^T$ and $\tilde K=\tilde Y (\tilde Y^T \tilde Y)^{-1} \tilde Y^T$. Let $H_{\lambda_X}=X(X^TX+\lambda_X \mathbf{I}_n)^{-1}X^T$, and $K_{\lambda_Y}=Y(Y^TY+\lambda_Y \mathbf{I}_n)^{-1}Y^T$. It is not difficult to show that $\tilde H=H_{\lambda_X}$ and $\tilde K=H_{\lambda_X}$. This observation provides an augmented-data interpretation of the Mantel test based on the ridge-penalized similarity matrices.

Note that the OLSE based on the augmented data depends only on $\lambda_X$, but not on $\lambda_Y$.  To examine the role played by $\lambda_Y$, we consider the augmented-data log-likelihood,
\begin{equation}
\label{eqn: neglik_augment}
l_{\lambda_X, \lambda_Y} = \mathbf{C} - (n+p+q)\log|\Sigma| - \text{tr}[(Y-X\beta)\Sigma^{-1}(Y-X\beta)^T]- \lambda_X\text{tr}[\beta\Sigma^{-1}\beta^T]-\lambda_Y \text{tr}[\Sigma^{-1}]
\end{equation}

\noindent where $\mathbf{C}$ is a constant that does not depend on any parameters. One interesting observation is that the likelihood corresponding to Equation \ref{eqn: neglik_augment} is mathematically identical to the posterior distribution of $\beta$ and $\Sigma$ where the joint prior is a normal-inverse-Wishart distribution:
\begin{eqnarray}
p(\beta,\Sigma) &\propto&  |\Sigma|^{\frac{p+q}{2}}exp\{-\frac{\lambda_X}{2}tr[\beta\Sigma^{-1}\beta^T]-\frac{\lambda_Y}{2} tr[\Sigma^{-1}]\}
\end{eqnarray}

\subsection{The Ridge as a Bridge Between Euclidean and Mahalanobis Metrics} \label{bridge_RV}
The Escoufier's RV coefficient \citealp{robert1976unifying} based on $H$ and $K$ is defined as
$$R(H,K)=\frac{\text{tr}(HK)}{\sqrt{\text{tr}(H^2)\text{tr}(K^2)}},$$
where $H$ and $K$ are similarity matrices, which are often p.s.d., and $H^2$ and $K^2$ represent the squares of $H$ and $K$, respectively, under the definition of matrix multiplication. Note that $R(H,K)$ is the sample Pearson's correlation of the vectorized similarity matrices $H$ and $K$. By examining the RV coefficients with different ridge penalty, we provide a novel unification of the fixed effects, ridge, and random effects model as a two-parameter class of correlation measures, parameterized by the ridge penalty terms $(\lambda_X,\lambda_Y) \in [0, \infty]^2$, where $\lambda_X=0$ corresponds to Euclidean similarity, and $\lambda_X=\infty$ corresponds to the Mahalanobis similarity.

\noindent \textbf{Theorem 1. (Limits of Multivariate Ridge Correlations.
)} \label{prop1} Let $X_{n\times p}$ and $Y_{n\times q}$ be column-centered data matrices of two sets of features, respectively. Let 
$H_{\lambda_x}=X(X^TX+\lambda_X\mathbf{I})^{-1}X^T$,  $K_{\lambda_y}=Y(Y^TY+\lambda_Y\mathbf{I})^{-1}Y^T$.
Then
\begin{equation}
   	\lim _{\lambda_X\rightarrow \lambda_X^*, \lambda_Y\rightarrow \lambda_Y^*} R(H_{\lambda_Y},K_{\lambda_X})=R(H_{\lambda_Y^*}, K_{\lambda_X^*}),
\end{equation}
where $H_0 = X(X^TX)^-X^T$, $K_0 = Y(Y^TY)^-Y^T$, $H_{\infty} = XX^T$, and $K_{\infty} = YY^T$.  

The proof is detailed in Appendix A.3 and outlined here. We first prove that $$\lim _{\lambda_X\rightarrow \lambda_X^*} R(H_{\lambda_X},K_{\lambda_Y})=R(H_{\lambda_X^*}, K_{\lambda_Y})$$
The result is obvious when $\lambda_X^*=0$. The proof for $\lambda_X^*=\infty$ is accomplished by performing singular value decomposition of $X$. Performing SVD of $Y$ and using a similar method will lead to the desired conclusion, i.e., 
$\lim _{\lambda_X\rightarrow \infty}\lim _{\lambda_Y\rightarrow \infty} R(H_{\lambda_X},K_{\lambda_Y})=R(H_\infty, K_\infty).$

This result indicates that, as we increase the ridge penalty, we are moving from Mahalanobis distances / similarities towards Euclidean distances / similarities. The quantities $R(H, K)$ and $tr(HK)$ are related to known statistics for many choices of $K$ and $H$.  For example, $tr(H_0K_0)$ equals the sum of the squared canonical correlations and is the Pillai’s trace statistic; in particular, it is multiple $R^2$ when $q=1$; it is also proportional to Hooper's trace correlation $R_T^2$ \citealp{hooper1959simultaneous}. Because $tr(H_\infty K_\infty)$ was derived from random effects model, after some derivations, it is not surprising to see that it is linked to method of moment estimates of variance components in the random effects model. 

To better understand the underlying geometric interpretation of different choices, we focus on the case of univariate outcomes, i.e., $q=1$ and assume that $Y$ has been standardized with mean zero and variance one. Note that the fixed and random effects models have the same null distribution of $Y$. Their corresponding score test statistics can be compactly written by considering the singular value decomposition (SVD) of $X$: $X = UDV^T$, where $U$ is $n \times r$ with $U^TU = I_r$, $D$ is an $r \times r$ diagonal matrix of singular values, $r=rank(X)$, $d_j, j = 1, \cdots, r$ are the non-zero singular values, and $V$ is $p \times r$, with $V^TV = I_r$.  We refer to the columns of $U$ as the \textit{principal directions} of $X$, and to $Z = U^TY$ as the \textit{principal correlation vector}, which is an $r \times 1$ vector with $j$th component $Z_j = R(Y, U_j)$, the Pearson correlation of $Y$ with $X$ along the $j$th principal direction. With these notations, $tr(H_0 YY^T)=\sum_{j=1}^r Z_j^2$ and $tr(H_\infty YY^T)=\sum d_j^2 Z_j^2$. Thus, we can interpret the fix effects score statistic as testing the Euclidean norm of the correlation vector; in contrast, the random effects (i.e., variance components) score statistic is equivalent to testing the weighted norm of $Z$, where the $j$th principal direction of $X$ is weighted by $d_j^2$, the variance of $X$ along the $j$th principal direction.  This has the effect of emphasizing the influence of directions in $X$ for which $X$ has large variance and reducing the influence of directions with small variance. The ridge regression score statistic equals $tr(H_{\lambda_X} YY^T)=\sum_{j=1}^r \frac{d_j^2}{d_j^2+\lambda_X} Z_j^2$, which is a compromise between the fixed effects and variance components tests, with small $\lambda_X$ yielding a test close to the fixed effects (and identical at $\lambda_X = 0$), and large $\lambda_X$ yielding a test close to the random effects score statistic, converging to identical tests as $\lambda_X \to \infty$.  In other words, the ridge statistic weights $Z_j^2$ proportional to $d_j^2$ as in the random effects statistic, but flattens each weight by a factor of $\frac{1}{d_j^2 + \lambda_X}$. This geometric interpretation motivates us to consider a ridge-penalized adaptive Mantel test that considers a set of ridge tuning parameters.

\section{Adaptive Mantel Test}
\label{adamant}
In this section, we develop the ridge-penalized adaptive Mantel test (AdaMant), a novel extension of the classical Mantel test to utilize the ridge penalty in association testing. A similar adaptive idea was proposed by Xu et al.\citealp{xu2017adaptive}. By using a permutation procedure on the minimum $p$-value from a set of test statistics, AdaMant is able to simultaneously test across a set of tuning parameters and kernels without increasing the type I error rate. In comparison to the RV-test, which relies on a single metric, AdaMant tests across a set of metrics, and can therefore achieve higher power, as shown by the simulations in Section \ref{simulations}.

\begin{figure}[ht]
\centering
\includegraphics[scale=0.6]{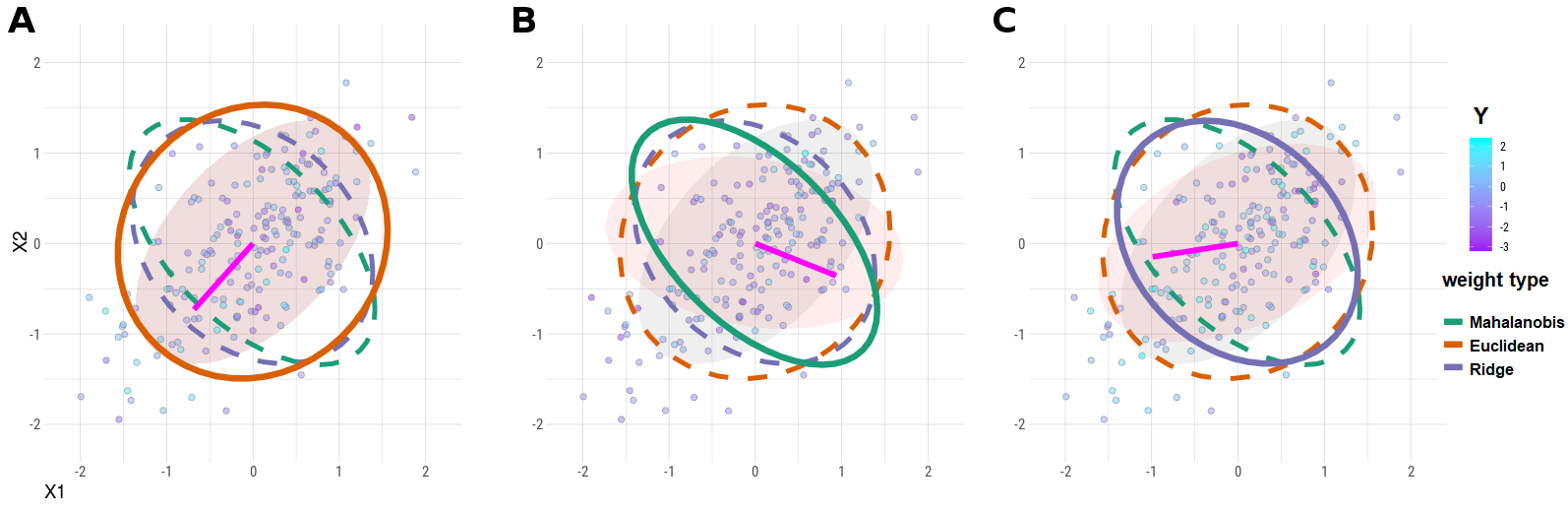}
\caption{Illustration of AdaMant for $Y$ generated from covariates $X$ rotated by angle $\theta$.  Gray shaded ellipses show the true covariance of $X$; pink shaded ellipses show the rotated covariances used to generate $Y$; magenta line segments show the direction of association estimated from a fixed effects model.  The solid lined ellipses correspond to the weights chosen by AdaMant; dashed ellipses are other weightings considered.  In each case, AdaMant selects the weighting that most emphasizes the empirical direction of association of $X$ and $Y$. (A) $\theta = 0$; (B) $\theta = 0.12 \pi$; (C) $\theta = 0.3 \pi$.}
\label{fig:adamant_three_plots}
\end{figure}

The idea of AdaMant can be understood through an example varying the relationship of the PCs of $X$ with $Y$ generated from a variance components model.  For the ease of presentation of the idea, we assume $Y$ is univariate, which implies that $\lambda_Y$ is not needed. Suppose that $X$ is a $100 \times 2$ matrix of covariates drawn iid from a standard normal distribution, and $Y \sim N(0, \sigma_b^2 K^{(\theta)} + \sigma^2\mathbf{I}_{100})$, with $K^{(\theta)} = U\Theta \Delta^2 \Theta^T U^T$ for $\Theta$ the $2 \times 2$ rotation matrix for angle $\theta$.  Figure \ref{fig:adamant_three_plots} shows the results of applying AdaMant with weights for the Euclidean, Mahalanobis, and ridge kernels ($\lambda_X = 10$) for data generated with $\theta = 0, 0.12 \pi, 0.3\pi$.  In each case, AdaMant selects the weighting that most emphasizes the empirical direction of association of $X$ and $Y$. This clearly demonstrates the potential advantage of AdaMant if the direction of association of $X$ and $Y$ is different than the first PC of $X$.  In this case, AdaMant will likely be a more powerful test than methods such as the RV test, which corresponds to using only the Euclidean weighting.

\subsection{Algorithm for the Adaptive Mantel Test}
\label{algorithm}
The ``adaptive''procedure receives as input a list of pairs of metrics/kernels $\{(\mathcal{K}^{\textbf{X}}_{m}, \mathcal{K}^{\textbf{Y}}_{m}) | m = 1 , \cdots M\}$ from which the matrices $K_m = \mathcal{K}^{\textbf{X}}_{m}(X)$ and $H_m = \mathcal{K}^{\textbf{Y}}_{m}(Y)$ are computed for each metric pair, $m = 1, \cdots, M$.  These metrics may be from a single family with varying tuning parameters, such as the ridge penalty tuning parameter, or may include kernels from different families. For each $m = 1, \cdots, M$, $P_m$ is calculated as the $P$-value of the Mantel test with metrics $\mathcal{K}_m^{\textbf{X}}$ and $\mathcal{K}_m^{\textbf{Y}}$ for $X$ and $Y$ respectively.  The AdaMant test statistic is defined as the minimum of these values, $P^{(0)} := \min_{m \in \{1, \cdots, M\}} P_m.$
A permutation procedure is practical for calculating the reference distribution for $P^{(0)}$.  For each $m$, and  $b = 1, \cdots, B$, $H_m^{(b)}$ is generated by permuting rows and columns of $H$ simultaneously, and the corresponding test statistic $P^{(b)}$ is calculated.  The AdaMant $P$-value is then calculated as
$$P_{AdaMant} = \frac{1}{B + 1} \sum_{b = 0}^B I\left(P^{(0)} \leq P^{(b)}\right).$$
\noindent General pseudocode for AdaMant is given in \textbf{Algorithm 1}. In this article, we focus on ridge kernels, which include Euclidean and Mahalanobis similarities as special cases. A discussion of the computational strategies and complexity for ridge kernels is given in Appendix A.4.

\begin{algorithm}
\caption{Adaptive Mantel Test}\label{alg:adamant}
\begin{algorithmic}[1]
\Procedure{AdaMant}{$X, Y, \{\mathcal{K}_m^{\textbf{X}}, \mathcal{K}_m^{\textbf{Y}}\}_{m = 1}^{M}$, B}
\State $H_m \gets \mathcal{K}^{\textbf{X}}_{m}(X),~~~m = 1, \dots, M$
\State $K_m \gets \mathcal{K}^{\textbf{Y}}_{m}(Y),~~~m = 1, \dots, M$
\State Calculate $T_{m}^{(0)} \gets T_{m} := \text{tr}(H_mK_m)$
\State Generate $B$ permutations of $H_m$, labeled $H_m^{(b)},~~~m = 1, \dots, M; b = 1, \dots, B$.
\State $T_{m}^{(b)} \gets \text{tr}(H_{m}^{(b)}K_m),~~~m = 1, \dots, M; b = 1, \dots, B$
\State $P_{m}^{(b)} \gets \frac{1}{B + 1}\sum_{b' = 0}^B I\left(Z_{m}^{(b)} \leq Z_{m}^{(b')}\right),~~~m = 1, \dots, M; b = 1, \dots, B$ 
\State $P^{(b)} \gets \min_{m = 1, \cdots, M} P_{m}^{(b)},~~~b = 1, \dots, B$
\State $P_{AdaMant} \gets \frac{1}{B + 1} \sum_{b = 0}^B I\left(P^{(0)} \leq P^{(b)}\right)$
\EndProcedure
\end{algorithmic}
\label{alg:alg1}
\end{algorithm}

\textbf{Remark.} Algorithm \ref{alg:alg1} computes the AdaMant $P$-value via a permutation procedure.  This method can be computationally expensive, but is likely feasible in practice when some computational tricks are used (see Appendix A.4). It is also possible to compute the asymptotic null distribution of the AdaMant test statistic using a moment-matching normal approximation and results on the order statistics of normal random variables \citealp{afonja1972moments}. However, this normal approximation for a mixture of chi-squared random variables is not accurate for small sample sizes \citealp{zhang2005approximate}, and so the permutation procedure is generally recommended.\\ 

\subsection{Choosing the Ridge Penalty}
\label{ridge-penalty}
A main advantage of the ridge-penalized AdaMant is to allow for simultaneous testing over a set of ridge tuning parameter values while incurring substantially less loss of power compared to multiple testing adjustment methods such as Bonferroni.  When applying AdaMant, the set of candidate penalty terms should be chosen as small as possible, since the test power will decrease as the number of penalty terms increases. When only ridge kernels of form $X(X^TX+\lambda_X \mathbf{I})^{-1}X^T$ are included in AdaMant, previous results on the role of the ridge penalty term in predictive modeling can help with the identification of a reasonable set of candidate values \cite{de2013prediction}. 

Specifically, it has been shown that, for column-standardized $X$, when the ridge penalty $\lambda_X$ is chosen to be proportional to the noise-to-signal ratio $\sigma^2/\sigma_b^2$, the resulting shrinkage estimator $\hat\beta_{\lambda_X}$ is identical to the best linear unbiased predictor (BLUP) for the random effects model. Recall that the genetic heritability is defined as $h^2=\frac{p\sigma_b^2}{\sigma^2+p\sigma_b^2}$, which implies that $\sigma^2/\sigma_b^2= p\frac{1 - h^2} {h^2}$.  Moreover, for a new observation with unknown response the predictions using the ridge and random effects models are the same when replacing $(H_\infty /p)^{-1}$ with $(H_\infty /[p\frac{1-h^2}{h^2}]+\mathbf{I})^{-1}$  (\citealp{de2013prediction}), which is equivalent to choose $\lambda_X = p \frac{1-h^2}{h^2}$. Note that, with fixed heritability, the choice of $\lambda_X$ should increase with $p$, which is not surprising as shrinkage/penalization is usually advantageous in high dimensional settings. Thus, a reasonable set of penalty terms can be determined in practice by positing \textit{a priori} a likely range for $h^2$, the scientific interpretation and range of plausible values of which depend on the specific modalities of $X$ and $Y$. For instance, for a common disease, $h^2 > 0.5$ is considered high (and would likely have already been discovered by previous studies), whereas $h^2 < 0.1$ is probably not scientifically interesting.  As a point of reference, most estimated heritability in the UK Biobank data is between 0.1 and 0.4  \citealp{ge2017phenome}.

\section{Simulations}
\label{simulations}

\subsection{Univariate Simulations ($q=1$)}
Because AdaMant is a permutation based test, it naturally controls its type I error rate at any desired level. Using simulated data generated from varying numbers of covariates and several covariance structures (homoskedastic, heteroskedastic, and compound symmetric), we confirmed that the type I error rate  was controlled appropriately. To assess the power of AdaMant, we first conducted simulations with $q=1$, $n = 200$, $p$ ranging from 50 to 500, $\sigma^2 = 1$, $\sigma_b^2=0.035^2$ (random effects), $\beta_j = (-1)^j 0.05, j = 1, \dots, p$ (fixed effects), and 500 permutations. The design matrix $X$ was generated from the matrix normal $N(0, \mathbf{I}_n, \Sigma_X)$, where $\Sigma_X$ is chosen to have a compound symmetric structure with $\rho =0.1$. For each setting, 500 simulations were run to estimate power. AdaMant was conducted using our \textit{adamant} R-package with $K = YY^T$ and $H_{\lambda_X}$ for $\lambda_X \in \{100, 1000, 2500, 5000, 7500, 25000), \infty\}$). Three other testing methods (aSPC \cite{xu2017adaptive}, dCOV \cite{szekely2007measuring}, and RV-coefficient) were included for comparison. 

\begin{figure}[ht]
\centering
\includegraphics[scale = 0.4]{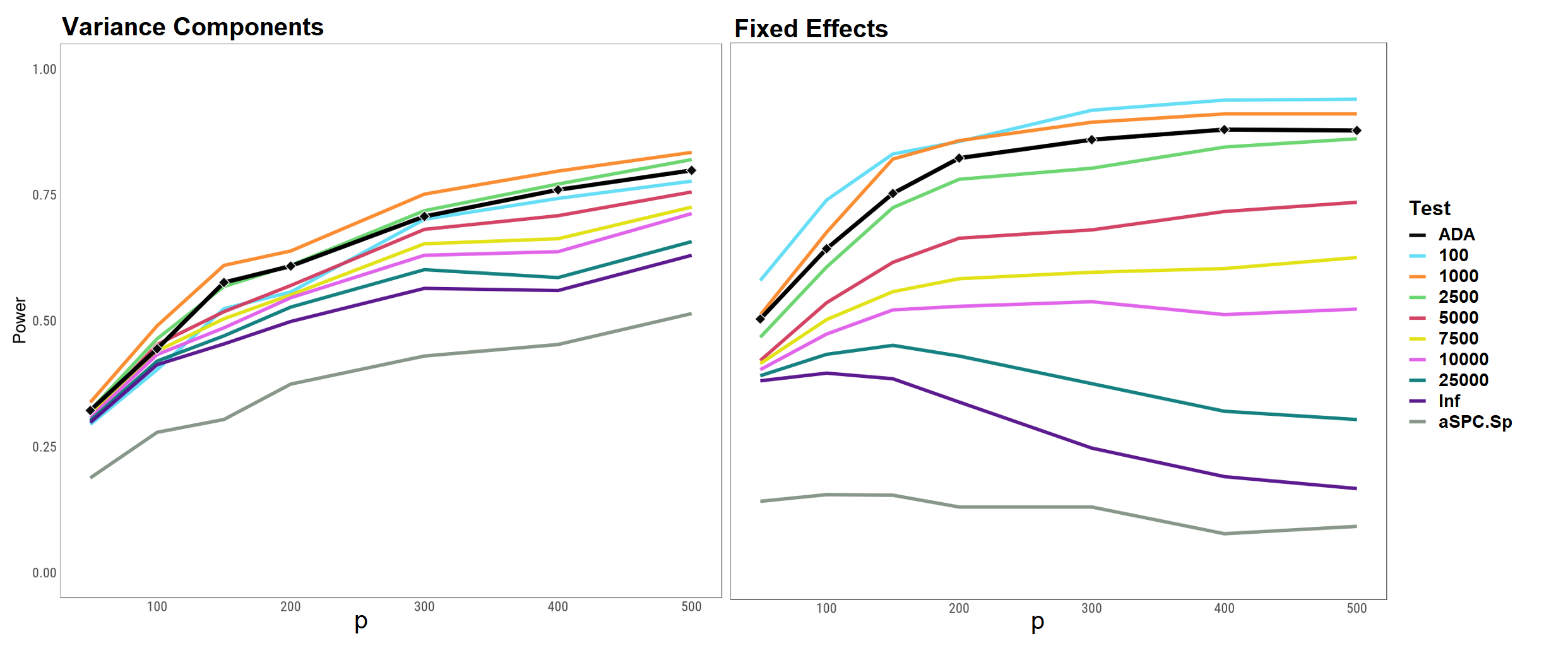}
\caption{Simulation results of the adaptive Mantel test.  Each setting used $n = 200$ observations, $\lambda_X \in \{100, 10^3, 2.5 \times 10^3, 5 \times 10^3, 7.5 \times 10^3, 10^4, 2.5 \times 10^4, \infty\}$, 500 test permutations, and 500 reps per value of $p$.  Error variance was fixed at $\sigma^2 = 1.$ The black curve is the adaptive Mantel power; the other curves are the power for the simple Mantel test with the ridge kernel with indicated penalty term, and the aSPC test.  The dCov and RV-coefficient tests perform nearly identical to the $\lambda_X = \infty$ test.  \textbf{(i)} Results for data generated from the variance components model with constant effect size $\sigma_b^2 = 0.035^2$ for each included feature. \textbf{(ii)} Power for data generated from a fixed effects model with $\beta_j = (-1)^j 0.05$.}
\label{fig:sims}
\end{figure}

Figure \ref{fig:sims} shows the estimated power for data generated from the variance components model (the left panel) and the fixed effects model (the right panel). The simulation results indicate that the power of AdaMant is competitive with the best of the single-parameter Mantel test for the values considered.  For the variance components setting, $\lambda_X = 1000$ attains the highest power for the ridge Mantel test.  As anticipated, the power of the ridge Mantel test exhibits unimodal behavior, such that the power increases to its peak for $\lambda_X = 1000$, and decreases as $\lambda_X \to \infty$.  We note that even though the variance components model is the true data generating mechanism, the test power is substantially increased over the variance components score test through the use of the ridge kernels.  The optimal ridge parameter selected by AdaMant ($\lambda_X = 1000$) confirms the relationship of the optimal parameter and the signal-to-noise ratio discussed \ref{ridge-penalty}, which predicts the optimal parameter is $\lambda^* = 1/0.035^2\approx 800$.

When the response data is generated from the fixed effects model, the simulation results show that using a penalized ridge kernel produces much higher power than competing tests, with $\lambda_X = 100$ giving the highest power across all values of $p$.  These results suggest that the variance components test is severely underpowered when the relationship of $X$ and $Y$ is strongest along principal components of $X$ with low variance.  Moreover, the RV coefficient, dCov, and aSPC tests perform equally poorly in this setting.

\subsection{Multivariate Simulations}
\subsubsection{Simulations with the Multivariate Variance Components Model}

 To compare the performance of AdaMant with existing methods for testing association of two multivariate feature sets, we generate data from multivariate versions of the variance components.  In the variance components simulation, $Y = G + \varepsilon$, with $Y \sim N(0, \Sigma_A \otimes XX^T / p + I_q \otimes I_n)$, where $\Sigma_A$ is a compound symmetric covariance matrix with diagonal entries equal to $\sigma^2_A$ and off-diagonal entries $0.1\sigma^2_A$.  We present simulation results for $n = 200, p = 40, q = 20$, and $\sigma^2_A \in \{0, 0.01, 0.02, 0.03, 0.04\}$.  We use ridge penalization for both $X$ and $Y$ features with $\lambda_X \in \{10, 100, \infty\}$ and $\lambda_Y \in \{10, 100, 500, 1000, \infty\}$.  The results (Table \ref{tab:sim_mv}) show that AdaMant power is comparable to the RV-test and dCov tests, all of which are substantially more powerful than the classical MT and aSPC test.

\subsubsection{Simulated EEG Data (q=20)}

\begin{figure}[ht]
\centering
\includegraphics[scale = 0.6]{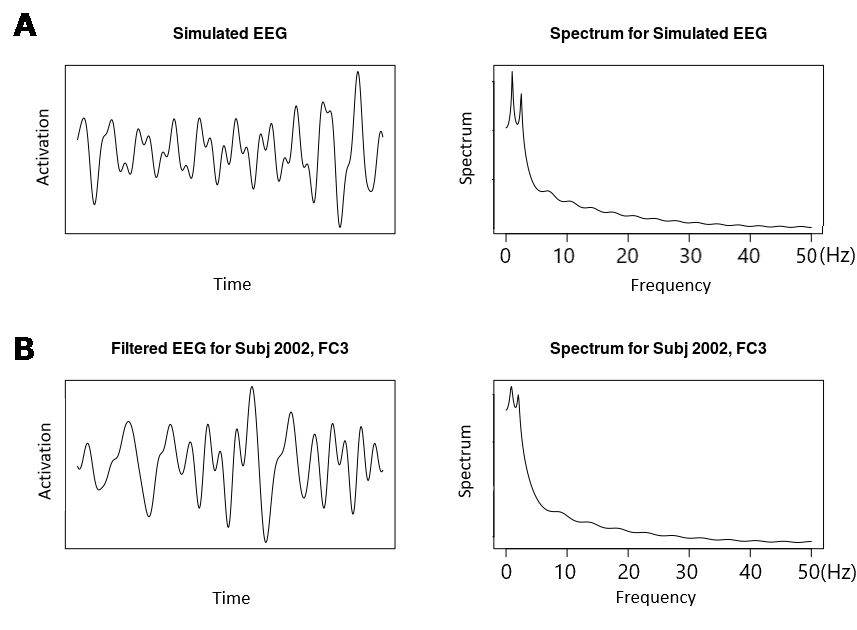}
\caption{A comparison of the simulated and real EEG time series shows the simulated series exhibit similar spectral characteristics as the real data after applying the same Butterworth bandpass filter (5 Hz - 50 Hz) (\textbf{Row A}) Filtered simulated EEG time series and corresponding spectrum. (\textbf{Row B}) Filtered EEG for subject 2002, channel FC4, and corresponding spectrum.}
\label{fig:simulated_eeg}
\end{figure}

Since we are primarily interested in applying AdaMant to imaging genetics data, we also consider the test performance on simulated data with structure similar to realistic EEG and single nucleotide polymorphism (SNP) data.  Simulated EEG data was produced as a mixture of AR(2) components \citealp{xu2019ess} to the ensure the spectral properties of the generated data are similar to real EEG data (Figure \ref{fig:simulated_eeg}). Association of the EEG data with the simulated genetic data was induced by generating the mixture weights with a variance components model with the genetic features as covariates. Similar to our real data analysis, the outcome of interest in this simulated EEG study is EEG coherence, where the coherence of two signals is a measure of the spectral similarity of two signals within a specified frequency band \cite{shumway2011time, ombao2008evolutionary}. Because the coherence matrix is symmetric, we use its vectorized upper triangles as the outcome variables. Further details on the simulation method and calculating of EEG coherence are given in Appendices \textbf{A.5, A.6}.


In applying AdaMant, similarity of observations in the EEG domain was computed from the ridge kernel applied to theta coherence matrices (resulting in 190 features); similarity in the genetics domain was computed from the ridge kernel applied to the observations' SNP vectors. Candidate penalty terms were $\lambda_X = \lambda_Y = \{10, 100, \infty\}$.  500 simulations were run for effect sizes equal to $\sigma_g^2 = \{0, 2.5 \times 10^{-6}, 5 \times 10^{-6}, 7.5 \times 10^{-6}, 1 \times 10^{-5}\}$.  The simulation results show (Figure \ref{fig:mv_sim_eeg}) that AdaMant has power competitive with other association testing methods in the two ways used for generating multi-channel EEG data.

\section{Applications}
\label{application}

We next consider data from 350 healthy college students from Beijing Normal University (BNU) who participated in a visual working memory task conducted by our co-author, during which 64-channel EEG was recorded at 1 kHz.  These data are part of a broader set of imaging, genetics, and behavioral data collected with the goal of identifying neurological features of brain connectivity that are significantly associated with both genetic features and cognitive performance.  In the present study, we focus on testing the association of EEG coherence with a set of genome-wide SNPs, and a set of candidate SNPs identified in a meta-analysis of educational attainment (EA). EEG coherence has been implicated in many cognitive processes and neurological diseases, for instance, in associative learning \citealp{fiecas2016modeling, ombao2018statistical, gorrostieta2019time}, as a predictor of ability in certain linguistic and visuospatial tasks \citealp{kang2017difference, park2014estimating}, and in Alzheimer's disease \citealp{engels2015declining, vecchio2018sustainable} and schizophrenia \citealp{griesmayr2014eeg}.

The experimental task is depicted in Figure \ref{fig:vwm_experiment_design_new}. The task for the participants was to recall the positioning of the red bars (the targets) on the left or right side of the image as indicated by the arrow in the center of the image.  Six experimental conditions were used for 2, 3, 4, 6, and 8 targets, and for 2 targets + 2 blue distractors.  Further details on experimental design and pre-processing of the EEG data are given in Appendix A.7.  The pairwise coherence for $q$ EEG channels is a $q \times q$ symmetric matrix, from which we extract the upper triangle and vectorize to form the $n \times \genfrac{(}{)}{0pt}{}{q}{2}$ matrix $Y$.  For our analysis, we consider coherence restricted to the 25 most anterior channels (``Frontal''), and fronto-parietal (FP) coherence from 30 channels covering the frontal and parietal lobes. This results in 300 distinct features for the 25 frontal channels, and 435 features for the fronto-parietal channels. Channel topography and coordinates are available from the BioSemi website.

Genotype data from approximately $5 \times 10^5$ single nucleotide polymorphisms (SNPs) were measured, of which 484,496 autosomal SNPs meeting quality control thresholds (minor allele frequency (MAF) $ > 1\%$, Hardy-Weinberg equilibrium (HWE) p-value $\ge 0.001$) were included for association testing.  Data processing was conducted using PLINK v1.90b4.4 and GCTA 1.91.5 beta2.  Let $X$ be the column-standardized SNP data matrix. Note that the purpose of association testing is to determine whether the heritability of EEG coherence in the theta (4 -- 8 Hz), alpha (8 -- 12 Hz), beta (12 -- 30 Hz), and gamma (30 -- 50 Hz) bands is significantly greater than zero. AdaMant was performed with $H_\infty = XX^T$ (which is proportional to the usual genetic relationship matrix), $K_{\lambda_Y} = Y(Y^TY + \lambda_Y I)^{-1}Y^T$ for $\lambda_Y \in \{10, 100, 1000, \infty\}$, using 5000 permutations.

\subsection{Association of EEG Coherence and All SNPs}
As mentioned above, we are interested in the coherence in the four bands (theta, alpha, beta, and gamma) at the Frontal region (300 coherence features from 25 channels) and the FP region (435 coherence features from 30 channels). We tested the association between each of the eight region-band combinations with all SNPs. Among them, the most significantly heritable coherence was found to be the theta band for the frontal channels, with $\mbox{p-value} = 0.0458$.  
Note that the p-value only provides significance but not magnitude or strength of the total genetic effect. We therefore estimated the multivariate genetic heritability using the method of moment estimator (see \citealp{ge2016multidimensional}). The estimated mean heritability of theta coherence for the frontal channels is $\hat h^2 = 0.165$. These evidences are at best marginal or weak, which is probably due to the small sample size, a potentially large number SNPs that do not contribute to EEG coherence, and low signal-to-noise ratio in EEG. Next, we focus on a subset of SNPs that are known to be associated with educational attainment.

\subsection{Association of EEG Coherence and SNPs Related to Educational Attainment}

SNPs were selected from a meta-analysis of educational attainment (EA) by \cite{okbay2016genome}, which identified over $3 \times 10^5$ SNPs with p-value less than 0.01, from which the top 100 EA SNPs were selected for testing.  AdaMant test was performed with $\lambda_Y \in \{0.5, 1, 5, 10, 100, 1000, \infty\}$ for ridge kernel similarity of the coherence data, and using 5000 permutations.  Genetic similarity of subjects was calculated as the $L_2$ inner product of the standardized SNP data.  Testing the association of EEG coherence of the Frontal and Fronto-parietal channels with the selected SNPs, we found that similarities from beta and gamma (30-50Hz) band coherence were significantly associated with the genetic features.  The beta band was the most significantly associated, with $\mbox{p-value} = 0.0104$ for the frontal channels and $P = 0.013$ for the FP channels. These results are consistent with existing evidence on the roles that the beta and gamma bands play in working memory \cite{lisman2013theta, proskovec2018beta}.  
\section{Discussion}
\label{discussion}

We have developed a unified framework of linear models that links the association measurements and score tests of the random effects (variance components), fixed effects, and ridge regression models as a single class of tests indexed by ridge penalty. 
The unified framework we discovered motivated us to develop the ridge-penalized adaptive Mantel test as a metric-based testing procedure, which reduces the difficulty of kernel parameter selection by simultaneously testing across a set of candidate parameters and automatically selecting the parameter yielding the most significant test from the candidate set.  As a tool for high-dimensional inference, AdaMant improves over competing association tests by data-adaptively selecting the kernel that yields the most significant association of $X$ and $Y$, without the need to explicitly adjust the calculated $P$-value.  AdaMant is simple to implement, yet can be usefully applied for any data modality that admits a meaningful measure of similarity.  As part of the present contributions, and to facilitate the use of AdaMant by other researchers, an R-package providing a basic implementation of AdaMant is available on the author's website.

AdaMant assesses the overall significance of association between two sets of variables. For example, in our application of AdaMant to the imaging genetics problem in Section 5.2, we examined the association between a set of brain signal variables defined by EEG coherence and a set of genetic variables based on SNPs. It is a global test thus exploratory. Once such overall significance is established, it is perhaps useful to further examine the contribution of individual variables. In Choi et al. \cite{choi2019contribution}, they developed tools to visualize the contribution of individual genetic variants to a set of variables for the structural changes in the brain measured from patients with Alzheimer’s disease. The effect they aimed to quantify is related to pleiotropy, the situation when one gene is associated with multiple phenotypes \cite{stearns2010one}. In many other situation, it might be interesting to estimate the overall genetic effects, i.e., genetic hertibility, for one specific or multiple targeting phenotypes. Whether one should examine the individual effects of $X$ on all/subsets of $Y$ or the individual effects of $Y$ on all/subsets of $X$ depends on the study goal and the desired direction of interpretation. Future research that adopts shrinking strategies, such as the ridge shrinkage considered here, to better understand the detailed relationship between two sets of variables might be helpful.

The present paper has primarily focused on the properties of AdaMant when only ridge kernel similarities are used, but this class of weights may not be optimal.  For instance, in Figure \ref{fig:adamant_three_plots} (B-C), weights that are concentrated along the true direction of association will produce a more powerful test than the ridge kernel weights.  Considering a wider class of similarity metrics, such as rotations of the Mahalanobis metric or non-linear kernel functions, is an important direction of future work for further improving the power and efficiency of high-dimensional testing methods. Many currently used association testing methods are powered for only a specific functional association of the two feature sets, e.g. linear or quadratic, or for general dependence testing (as with dCov).  The idea of using a ridge penalty in defining distance is flexible and can be incorporated into other powerful tests. For example, the dCov test with the Euclidean distance is able to capture both linear and nonlinear relationship as a zero dCov implies statistical dependence of the two feature sets \citealp{szekely2007measuring}. One way might improve the dCov test is to replace the Euclidean distance in dCov with a ridge distance indexed by a penalty parameter $\lambda$. AdaMant using the Gaussian radial basis kernel with a suitable selection of tuning parameters may be able to detect a wide variety of associations (although not all).  Building on kernel-based methods for estimating heritability \citealp{liu2007semiparametric}, future work will consider extensions of AdaMant to a more general class of similarity measures, and investigate methods for improving testing power while selecting across a large number of tuning parameters.

The primary goal of this article is to understand the role that ridge penalization plays in similarity or distance measurements and to develop a ridge-penalized adaptive Mantel test. We notice that in the past few years, with the counterintuitive discovery that overfitting often leads to accurate predictive models in deep learning \cite{zhang2016understanding}, there has been increased interest in investigating overfitting for estimation and prediction when $p\gg n$. More recently, it has been found that the optimal ridge penalty is positive when the coefficients are generated from a distribution \cite{hastie2019surprises} whereas the optimal ridge penalty could be negative when the coefficients are fixed \cite{kobak2018optimal}. These studies rely on random matrix theorems and assumptions of the covariance structure of the explanatory variable. Although our goal is different, our simulation results nevertheless are consistent with theirs. As shown in Figure 2, ridge penalization is more helpful when data were generated using a random effects model and less helpful when data were generated from a fixed effects model. In practice, the true mechanism is never known; thus it is perhaps reasonable to be data driven, i.e. let the data determine the optimal amount of penalization. We will continue examining whether and when mitigating overfitting by ridge penalty is beneficial for hypothesis testing.

\appendix

\section*{\textbf{A.1 Relationship between Squared Distance and Similarity}}
As assumed in the paper, both $X_{n\times p}$ and $Y_{n\times q}$ have been column centered. The squared weighted distance between subjects $i$ and $j$ is 
$$D_{\mathbf{W},ij}^2=(x_i-x_j)^T\mathbf{W}(x_i-x_j)=x_i^T\mathbf{W}x_i-2x_i^T\mathbf{W}x_j+x_j^T\mathbf{W}x_j,$$
where $x_i^T\in \mathbf{R}^p$ is the $i$th row of the data matrix $X$. 

Let $C$ be the ``centering" matrix, i.e., $C=\mathbf{I}-\frac{1}{n}\mathbf{1}\mathbf{1}^T$. Let $e_i$ denote the vector whose $i$th element and 1 and the other elements are 0. Then the $(i,j)th$ element of $-\frac{1}{2}CD_\mathbf{W}^2C$ is 
$-\frac{1}{2}(e_i-\mathbf{1}/n)^T D_{\mathbf{W}}^2 (e_j-\mathbf{1}/n)$. Note that

\begin{align*}
(e_i-\mathbf{1}/n)^T D_\mathbf{W}^2 (e_j-\mathbf{1}/n) &= e_i^TD_W^2e_j - e_i^TD_\mathbf{W}^2\mathbf{1}/n - \mathbf{1}^T D_\mathbf{W}^2e_j/n + \mathbf{1}^TD_\mathbf{W}^2\mathbf{1}/n^2\\
&= D_{\mathbf{W},ij}^2 - \frac{1}{n} \sum_{j} D_{\mathbf{W},ji}^2 -\frac{1}{n} \sum_j D_{\mathbf{W},ij}^2 + \frac{1}{n^2}\sum_{i,j} D_{\mathbf{W},ij}^2\\
&=  x_i^T\mathbf{W}x_i -2x_i^T\mathbf{W}x_j + x_j^T\mathbf{W}x_j   \\
&~ -\frac{2}{n} \sum_j (x_i^T\mathbf{W}x_i - 2x_i^T\mathbf{W}x_j + x_j^T\mathbf{W}x_j) \\
&~ +\frac{1}{n^2} \sum_{ij}(x_i^T\mathbf{W}x_i - 2x_i^T\mathbf{W}x_j + x_j^T\mathbf{W}x_j) \\
&= -2x_i^T\mathbf{W}x_j -\frac{2}{n}x_i^T\mathbf{W}\sum_j x_j + \frac{1}{n^2}\sum_i(x_i^T\mathbf{W}\sum_jx_j)\\
&=-2x_i^T\mathbf{W}x_j
\end{align*}
    
The last step is true because the columns of $X$ have been centered. As a result, the $(i,j)$th element of $-\frac{1}{2}CD_\mathbf{W}^2C$ is $x_i^T\mathbf{W}X_j$, which is the $(i,j)$th element of $X\mathbf{W}X^T$. 

Note that
\begin{align*}
    & \mbox{Mahalanobis: } \mathbf{W}=(X^TX)^- \Rightarrow X\mathbf{W}^{-1}X^T=H_0 \\
    &\mbox{Ridge penalized: }  \mathbf{W}=(X^TX+\lambda_{X}\mathbf{I})^{-1} \Rightarrow X\mathbf{W}X^T=H_{\lambda_X}\\
    &\mbox{Euclidean: } \mathbf{W}=\mathbf{I} \Rightarrow X\mathbf{W}^{-1}X^T=H_\infty 
\end{align*}
This established the relationship between square distance and similarity (weighted inner product).

\section*{\textbf{A.2 Proof of Proposition 1}}
\noindent \textbf{Proposition 1}. Let $H_0=X(X^TX)^{-1}X^T$, $K_0=Y(Y^TY)^{-1}Y^T$, $H_\infty=XX^T$, and $K_\infty=YY^T$. The score statistic from the fixed effects model is proportional to $tr(H_0K_0)$.  When $\Sigma = \sigma^2 \mathbf{I}_q$ and $\Sigma_b=\sigma_b^2 \mathbf{I}_q$, which is always true when $q=1$, the score statistic from the random effects model is proportional to $tr(H_\infty K_\infty)$.  Asymptotically, their null distributions are chi-squared and mixture of chi-squared, respectively. 

\noindent \textbf{Proof}. The derivations are mainly based on gradients with respect to vectors and matrices. See the Matrix Cook Book \cite{petersen2012matrix} for a collection of useful derivative results.

\begin{enumerate}
    \item[] \underline{Fixed effects model}. Suppose the random matrix $Y$ follows a matrix normal distribution: $Y\sim N(X\beta, \mathbf{I}_n, \Sigma)$. Equivalently, $vec(Y)\sim N(vec(X\beta), \Sigma \otimes \mathbf{I}_n)$, where $vec(\cdot)$ denotes the vectorization operator. The log-likelihood is 
    $$l=\mathbf{C}-\frac{1}{2} tr[\Sigma^{-1}(Y-X\beta)^T(Y-X\beta)]= \mathbf{\tilde C}-\frac{1}{2}tr[\Sigma^{-1}\beta^TXX^T\beta ]+tr[\Sigma^{-1}Y^TX\beta],$$

\noindent where $\mathbf C$ and $\tilde{\mathbf C}$ are constant terms that do not involve parameters of interest. 

The score is $U_\beta=\frac{\partial l}{\partial \beta}|_{\beta=0} = X^TY\Sigma^{-1}$. Under the null hypothesis $\beta=0$, it follows the matrix normal distribution $N(0, X^TX, \Sigma^{-1})$. Equivalently, we have $vec(U_\beta)\overset{H_0}\sim N(0, \Sigma^{-1} \otimes (X^TX))$. The corresponding score statistic is 
\begin{align*}
vec(U_\beta)^T[\Sigma^{-1} \otimes (X^TX)]^{-1}vec(U_\beta) &=
tr[X(X^TX)^{-1}X^TY\Sigma^{-1}Y^T]\\
&\overset{\cdot}= tr[X(X^TX)^{-1}X^T Y(Y^TY)^{-1}Y^T]/n
\end{align*}
In the last step we replaced $\Sigma$ with a consistent estimator $tr(Y^TY)/n)$. Thus, the score statistic for the fixed effects model is proportional to $tr(H_0K_0)$, where $H_0=X(X^TX)^{-1}X^T$ and $K_0=Y(Y^TY)^{-1}Y^T$. Note that $tr(H_0K_0)$ is Pillai's trace statistic, which also equals the sum of the squared canonical correlations.

    \item[] \underline{Random effects model}. Under the random effects model, $vec(Y)\sim N(0, V)$, and the log-likelihood is
        $$l=\mathbf{C}-\frac{1}{2}log|V|-\frac{1}{2} vec(Y)^TV^{-1}vec(Y)$$
where $V=\Sigma_b \otimes XX^T + \Sigma \otimes \mathbf{I}_n$. When $\Sigma_b=\sigma_b^2 \mathbf{I}_q$, the score is
$$U_{\sigma_b^2} = -\frac{1}{2}tr[\Sigma^{-1}]tr[XX^T]+\frac{1}{2}vec(Y)^T(\Sigma^{-2}\otimes XX^T)vec(Y)$$

Note that the first term in $U_{\sigma_b^2}$ is the expected value of the second term under the null hypothesis. Similar to others \cite{liu2007semiparametric, huang2013gene}, we use the quadratic part, which can be rewritten as $tr[XX^TY\Sigma^{-2}Y^T]$. When $\Sigma=\sigma^2\mathbf{I}_q$, up to a scaling factor, the score test statistic is equivalent to $tr[XX^TYY^T]=tr(H_\infty K_\infty)$. 

\underline{Large-sample distributions}. It is known that when the sample size $n$ is large and the null hypothesis is true, the distribution of a quadratic form can be approximated by a mixture of chi-square distributions. 

For $tr(H_0K_0)$, the weights are equal, as the non-zero eigenvalues of $H_0=X(X^TX)^{-1}X^T$ are all 1. Thus, after re-scaling $tr(H_0K_0)$ appropriately, its null distribution is approximately a chi-squared distribution when the null hypothesis is true. As a comparison, for  $tr(H_\infty K_\infty)$, the weights depends on the singular values of $X$, which is noted in earlier work.

\end{enumerate}

\section*{\textbf{A.3 Proof of Theorem 1}}

\noindent \textbf{Theorem 1. (Limits of Multivariate Ridge Correlations.
)} Let $X_{n\times p}$ and $Y_{n\times q}$ be column-centered data matrices of $p$ features and $q$ features, respectively. Let 
$$H_{\lambda_x}=X(X^TX+\lambda_X\mathbf{I})^{-1}X^T \mbox{ , } K_{\lambda_y}=Y(Y^TY+\lambda_Y \mathbf{I})^{-1}Y^T$$
Then
\begin{equation}
   	\lim _{\lambda_X\rightarrow \lambda_X^*, \lambda_Y\rightarrow \lambda_Y^*} R(H_{\lambda_Y},K_{\lambda_X})=R(H_{\lambda_Y^*}, K_{\lambda_X^*}),
\end{equation}
where $H_0 = X(X^TX)^-X^T$, $K_0 = Y(Y^TY)^-Y^T$, $H_{\infty} = XX^T$, $K_{\infty} = YY^T$.

\noindent \textbf{Proof}. Consider the singular value decomposition (SVD) of $X$.  Let $\text{rank}(X) = r$ where $r \leq \min(p, n)$, the SVD is $X = UDV^T$, where $U$ is $n \times r$ with $U^TU = I_r$, $D$ is an $r \times r$ diagonal matrix of singular  values $d_j, j = 1, \cdots, r$, and $V$ is $p \times r$, with $V^TV = I_r$. Let $Z=U^TY(Y^TY+\lambda_Y I)^{-1/2}$. We have

\begin{align*}
\lim _{\lambda_X\rightarrow \infty} R(H_{\lambda_X},K_{\lambda_Y})
    &=\frac{1}{\sqrt{tr(K_{\lambda_Y}^2)}} \lim _{\lambda_X\rightarrow \infty} \frac{tr(H_{\lambda_X} K_{\lambda_Y})}{tr(H_{\lambda_X}^2)}\\
    &= \frac{1}{\sqrt{tr(K_{\lambda_Y}^2)}} \lim _{\lambda_X\rightarrow \infty}
    \frac{tr(Z^T DV^T (X^TX+\lambda_X I)^{-1}V D Z)}{tr(UDV^T(X^TX+\lambda_X I)^{-1}VD^2V^T(X^TX+\lambda_X I)^{-1}VDU^T)}\\
    &= \frac{1}{\sqrt{tr(K_{\lambda_Y}^2)}} \lim _{\lambda_X\rightarrow \infty} \frac{\sum_{j=1}^r d_j^2Z_j^2/(d_j^2+\lambda_X)}{\sqrt{\sum_{k=1}^r \left(
    \frac{d_k^2}{d_k^2+\lambda_X}\right)^2}}\\
    &= \frac{1}{\sqrt{tr(K_{\lambda_Y}^2)}} \lim _{\lambda_X\rightarrow \infty} 
    \sum_{j=1}^r d_j^2Z_j^2 \left[(d_j^2+\lambda_X)^2\sum_{k=1}^r \left(
    \frac{d_k^2}{d_k^2+\lambda_X}\right)^2\right]^{-1/2}\\
    &= \frac{1}{\sqrt{tr(K_{\lambda_Y}^2)}}  
    \sum_{j=1}^r d_j^2Z_j^2 \left[\sum_{k=1}^rd_k^4 \lim _{\lambda_X\rightarrow \infty}\left(
    \frac{d_j^2+\lambda_X}{d_k^2+\lambda_X}\right)^2\right]^{-1/2}\\
    &=\frac{\sum_{j=1}^r d_j^2Z_j^2}{\sqrt{tr(K_{\lambda_Y}^2)} \sqrt{\sum_{k=1}^rd_k^4} }\\
    &=\frac{ tr(H_\infty K_{\lambda_Y})}{\sqrt{tr(H_\infty^2)}\sqrt{tr(K_{\lambda_Y}^2)} }\\
    &=R(H_\infty, K_{\lambda_Y})
\end{align*}

Similarly, we can show that 
\begin{align*}
\lim _{\lambda_Y\rightarrow \infty} R(H_{\infty},K_{\lambda_Y})&=R(H_\infty, K_\infty),
\end{align*}
and it follows that 
\begin{align*}
\lim _{\lambda_X\rightarrow \infty}\lim _{\lambda_Y\rightarrow \infty} R(H_{\lambda_X},K_{\lambda_Y})&=R(H_\infty, K_\infty),
\end{align*}

\section*{\textbf{A.4 Computational Complexity}}

If the feature space is very high-dimensional or if $n$ is large, a straightforward implementation of Algorithm 1 may be computationally impractical.  However, when only ridge kernels with varying values of $\lambda$ are included in AdaMant,  are two approaches that can be used to greatly reduce the computational cost.

The first approach utilizes the SVD $X = UDV^T$.  The computational complexity needed for finding the SVD for $X$ is $O(np^2)$. Once the SVD is computed, we can compute $Z=U^TY$ and $S_\lambda=\sum_{i=1}^r \frac{\eta_i}{\lambda+\eta_i}Z_i^2$, which has a total complexity of $O(nr)$. Note that when $n\gg p$, the rank $r$ is no greater than  $p$; as a result, the cost needed for calculating $S_\lambda$ is $O(nr)\approx O(np)$. Calculating the test statistics for $B$ permutations requires $O(Bn^2)$, for a total computational complexity of $O(np^2+ Bnp)$. 

On the other hand, when $p$ is very large relative to $n$, SVD of $X$ is not the best option. In this situation, we can use the identity \cite{henderson1981deriving} $X(X^TX + \lambda I)^{-1} = (XX^T + \lambda I)^{-1}X,$ so that the dimension of the matrix to be inverted is lowered to $n\times n$, rather than $p\times p$.  From this identity, $K_\lambda$ can be rewritten as 
$K_\lambda = (XX^T + \lambda I_n)^{-1}XX^T.$  Note that calculating $K_\lambda$ involves multiplying the $n\times p$ matrix $X$ and the $p\times n$ matrix $X^T$, multiplying two $n\times n$ matrices, and inverting an $n\times n$ matrix. When $p\gg n$, the computation cost is dominated by calculating $XX^T$, which has a complexity of $O(n^2p)$. The Mantel test statistic can be calculated as $tr(YY^T K_\lambda)=Y^T K_\lambda Y,$ which has a complexity of $O(n^2)$. With $B$ permutations, the total computational complexity is $O(n^2p + n^2B)$, which is less than the required computational complexity using SVD.

To generate the genetic data, observations were placed into one of two balanced groups to form two groups of genetically similar observations.  For groups $k = 1, 2$, a group-wide SNP vector $X^{(k)}$ of length 300 is generated by iid sampling from a discrete uniform distribution over $\{-1, 0, 1\}$; for subject $i$ in group $k$, a subject-specfic error vector $\varepsilon^{(k)}_{i} \sim N(0, I_{300})$ are added to the group SNP vector to get the subject SNP vector $X^{(k)}_i = X^{(k)} + \varepsilon^{(k)}_i$.  To generate data with spectral characteristics similar to real EEG data, for each observation 20 time series of length 1000 were generated (for 20 ``channels'') as a weighted mixture of two independent AR(2) processes with spectral peaks at 5.12 Hz and 12.8 Hz and second-order AR-coefficient fixed at -0.99, resulting in a time series with two peaks at the chosen frequencies, with the magnitude of the peaks determined by the mixing weights.  The mixing weights for a random selection of 10 of the channels (common across all subjects) are drawn iid from $N(0, 1)$, so that no genetic effect exists for these channels.  To link the genetic features to the remaining channels, the $200 \times 10 \times 2$ array $W$ of AR(2) mixing weights was generated from a multivariate variance components model, such that for channel $j$ and AR(2) basis $m$, the subject weight vector is generated from $W_{jm} \sim N(0, \sigma^2_b XX^T)$, where $\sigma^2_b$ is varied across simulations to control the strength of genetic association.  The pair of weights for each subject and channel are then each squared and divided so that the weights sum to one.  From the resulting EEG data, the connectivity matrix for each observation is calculated from the pairwise coherence for all pairs of channels.

\section*{\textbf{A.5 Simulating Realistic Imaging Genetics Data}}
\label{simulation_method}

We generated imaging and genetic data for $n=200$ subjects, each with $p=300$ SNPs and $q=20$ EEG channels. To generate the genetic data, observations were placed into one of two balanced groups to form two groups of genetically similar observations.  For groups $k = 1, 2$, a group-wide SNP row vector $X^{(k)}$ of length 300 is generated by iid sampling from a discrete uniform distribution over $\{-1, 0, 1\}$; for subject $i$, a subject-specific error row vector $\varepsilon_{i} \sim N(0, I_{300})$ is added to the group SNP vector to get the subject SNP vector $X^{(k)} + \varepsilon_i$, where $k$ is the group that subject $i$ belongs to. Finally, we stack the data together to create the $n\times p$ (=$200\times 300$) genetic data matrix $X$:
$$X=
\begin{pmatrix}
1 & 0\\ \cdots &\cdots \\ 1& 0\\
0 & 1\\ \cdots &\cdots \\ 0& 1
\end{pmatrix}
\begin{pmatrix}
X^{(1)} \\ X^{(2)}
\end{pmatrix} + \epsilon
$$

To generate data with spectral characteristics similar to real EEG data, for each observation 20 time series of length 1000 were generated (for 20 ``channels'') as a weighted mixture of two independent latent AR(2) processes with spectral peaks at 5.12 Hz and 12.8 Hz and second-order AR-coefficient fixed at $-0.99$, resulting in a time series with two peaks at the chosen frequencies, with the magnitude of the peaks determined by the mixing weights. Let $Z^{(1)}(t)$ and $Z^{(2)}(t)$ denote the two time series. The mixing weights for the first 10 channels (common across all subjects) are drawn iid from N(0,1), so that no genetic effect exists for these channels. We use two methods to generate AR mixing weights linking the genetic features to the remaining channels.  In each case, after generating the initial weights using the methods described below, the weights for each subject and channel are then squared and scaled so that the weights are nonnegative and sum to one.  From the resulting EEG data, the connectivity matrix for each observation is calculated from the pairwise coherence for all pairs of channels. 

\begin{figure}
\centering
\includegraphics[width=\textwidth]{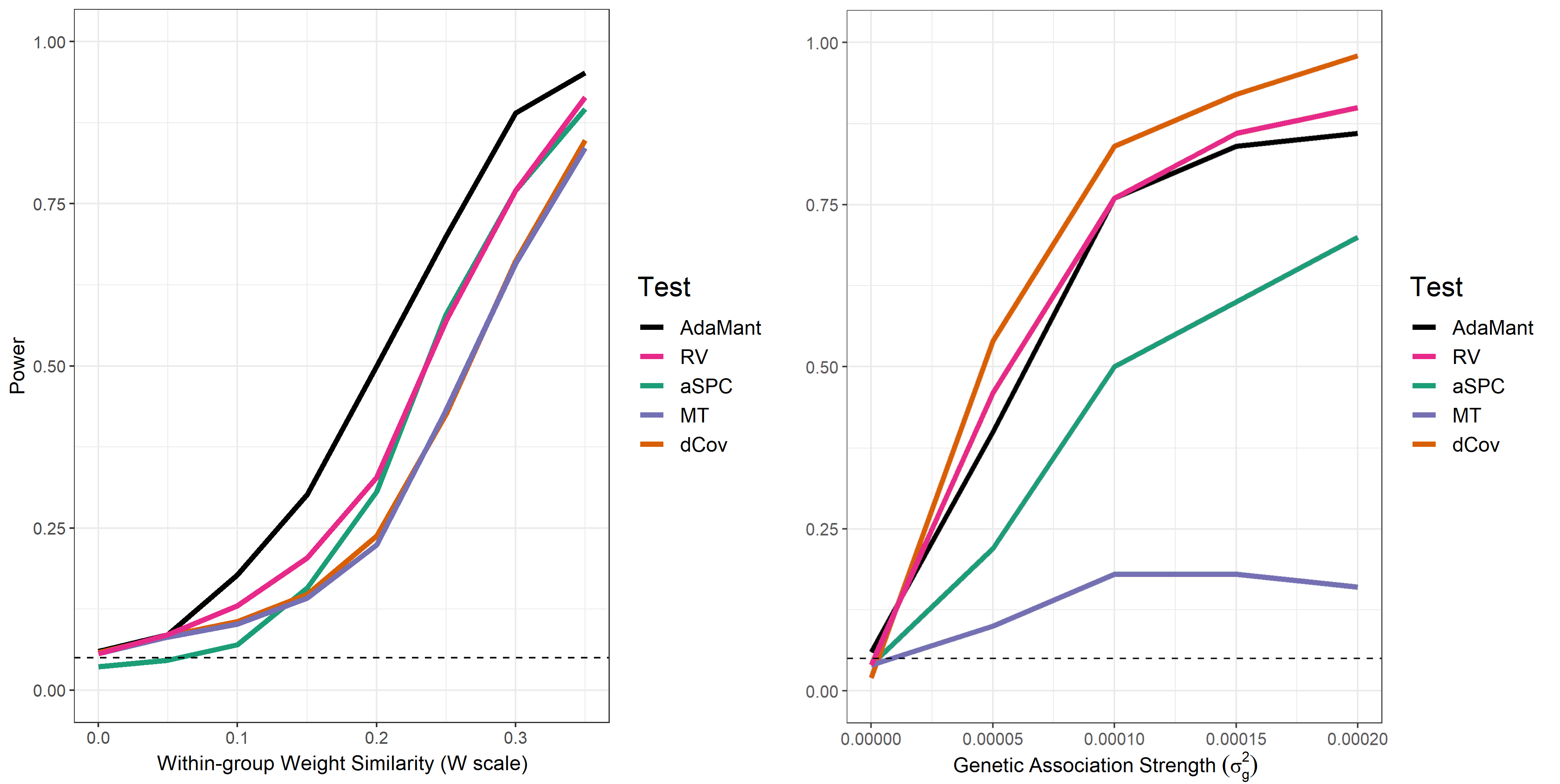}
\caption[Simulation results for multivariate adaptive Mantel test power]{Power plots of association testing methods for simulated 25-channel EEG data with $n = 200$ observations, $p = 300$ genetic features, $q = 20$ EEG channels. (\textit{left}) Simulation with strength of association controlled by the within-group genetic similarity, for $\sigma^2_g = 5\times 10^{-5}$ fixed. (\textit{right}) Simulation with strength of association controlled by $\sigma^2_g$.}
\label{fig:mv_sim_eeg}
\end{figure}

\begin{figure}[ht]
\centering
\includegraphics[scale = 0.6]{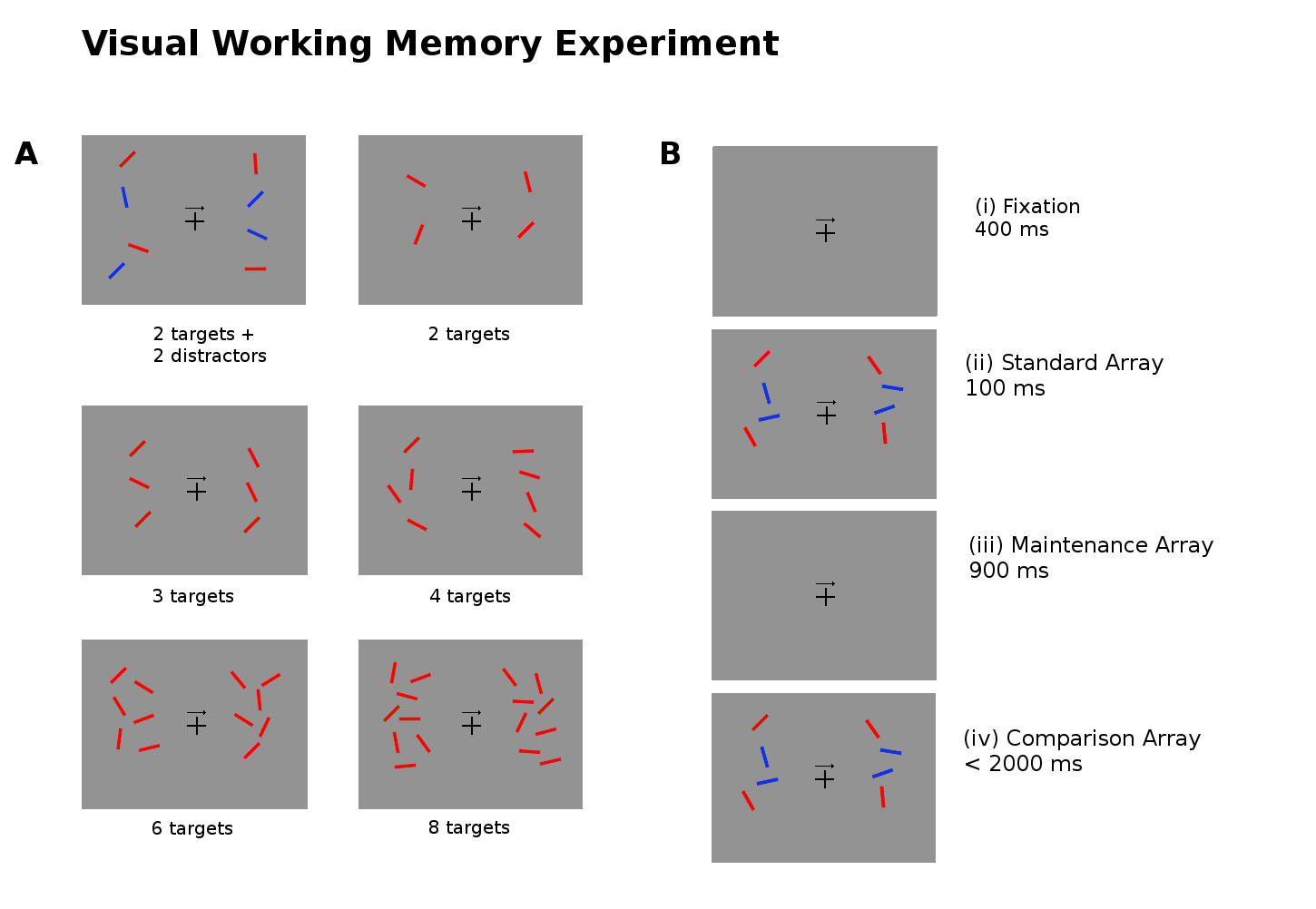}
\caption{(A) The experimental task required subjects to remember the positioning of the red bars (the targets) on the left or right side of the image as indicated by the arrow in the center of the image.  The six images show examples of the test image for different numbers of targets and distractors. (B) For each trial of the experiment, subjects are shown (i) a fixation image ($400 \pm 200$ ms); (ii) the standard array to be memorized (100 ms); (iii) the maintenance image (900 ms); (iv) the comparison array, at which point subjects are asked to respond within 2000 ms if the targets in the comparison arrays \textit{match} or \textit{mismatch} the standard array.}
\label{fig:vwm_experiment_design_new}
\end{figure}

In the first approach for generating mixing weights, a set of common weights is produced for each group $k$ of "genetically" similar subjects, where the weight for channel $j$, group $k$, and AR component $m$ is generated as  $W_{jkm} = W_{scale} \cdot W'_{jkm}$, for $W'_{jkm} \sim Bern(p)$. With this method, $W_{scale}$ controls the within-group similarity of the mixing weights, and $p \in [0, 1]$ controls the expected number of channels that have similar weights within each group; in the present simulations we set $p = 0.1$.  The mixing weight for subject $i$, channel $j$, group $k$, and component $m$ is then generated as $W_{jkm} + \omega_{ijmk}$, where $\omega_{ijkm}$ is an idiosyncratic positive perturbation drawn from a log-normal distribution with parameters $\mu_{\omega} = 0.5, \sigma_{\omega} = 0.5$.  The results for data generated with this approach as shown the left panel of Figure \ref{fig:mv_sim_eeg}.

In the second approach, we generate mixing weights from a multivariate variance components model so that genetically similar subjects have similar weights. Let $Y_j(t)$ denote the $200\times 1$ vector for the $j$th EGG channel of the 200 subjects. This can be compactly expressed as mixing of the two latent time series, which is specified by the following equation
$$Y_j(t)= W_{0j} I(1\le j\le 10) + 
\begin{pmatrix} W_{1j} |& W_{2j} \end{pmatrix}
\begin{pmatrix}Z^{(1)}(t)\\ Z^{(2)}(t)\end{pmatrix} I(11\le j\le 20)$$
where $W_{0j} \sim N(0, \mathbf{I_{200}})$ and $W_{1j}, W_{2j} \overset{iid} \sim N(0,\sigma_g^2 XX^T)$. 
For this approach, we can control the strength of association by increasing $\sigma^2_g$, which increases the likelihood that subjects with similar genetic covariates will have similar mixing weights (globally across both groups).  The results for data generated with this approach as shown the right panel of Figure \ref{fig:mv_sim_eeg}.

\section*{\textbf{A.6 Calculating Coherence}}
The coherence between two EEG channels at a particular frequency $\omega$ is a measure of the oscillatory concordance of the the two signals at $\omega$.  Coherence matrices for each subject were calculated as follows.

 \begin{enumerate}
   \item
   Suppose we have $Q$ channels and denote the time series at the $q$-th channel to be $X_q(t)$ where $t=1, \ldots, T$ \ and $q=1, \ldots, Q$.

  \item \textbf{Fourier transform}:
   We compute the discrete Fourier transform (DFT) 
   of the time series at each channel $q$ 
   \begin{center}
      $d_q(\omega_j)=T^{-1/2} \sum\limits^T_{t=1}X_q(t)\exp(-2\pi i\omega_j t)$
  \end{center}
   for $j=0,1,...T-1$, where the standardized 
   frequencies $\omega_j=j/T \in (-0.5, 0.5]$. Denote the DFT of the $q$-th channel at frequency $\omega_j$ to be $d(p,j)$. The cross spectrum of $(q_1, q_2)$-th channels at frequency $\omega_j$ is $d(q_1,j)\cdot d(q_2,j)^*$ (where $*$ denotes conjugate transpose). This is the $(q_1,q_2)$-th element of spectral matrix at frequency $\omega_j$.
   Denote the spectral matrix at frequency $\omega$ as\\

 \[
 S(\omega) = \begin{pmatrix}
       S_{11}(\omega)  & S_{12}(\omega) &  \cdots & S_{1Q}(\omega) \\
       S_{21}(\omega)  & S_{22}(\omega) &  \cdots & S_{2Q}(\omega) \\
       \cdots    & \cdots &  \cdots & \cdots   \\
       S_{Q1}(\omega)  & S_{Q2}(\omega) &  \cdots & S_{QQ}(\omega) \\
   \end{pmatrix}
 \]

   \item \textbf{Averaging over frequencies within bands}: We set the cutoff value for frequency to be 45 Hertz (which is equivalent to the standardized frequency $\omega=45/256$. For each band, we average the spectral matrix over frequencies within that band.

   \item \textbf{Averaging over trials}: We average the spectrum matrix for each band over 359 trials and get the 3-d array  ($channel\times channel\times bands$) which contains \textbf{spectral matrix} for each frequency band.
   \item \textbf{Calculate coherence matrix}: For each subject, We calculate the coherence matrix for each frequency band based on spectral matrices. The coherence between channel m and n can be calculated as
   \begin{equation*}
       r_{mn}(f)=\frac{|S_{mn}(f)|^2}{S_{mm}(f)S_{nn}(f)}
   \end{equation*}
 \end{enumerate}

\section*{\textbf {A.7 Design of EEG Working-memory Experiment}}
Each trial consists of the presentation of four images in sequence, with a total duration approximately three seconds, structured as follows: (i) \textit{fixation} image for $400 \pm 200$ ms; (ii) the \textit{standard array} to be memorized (100 ms); (iii) the blank \textit{maintenance} image (900 ms); (iv) \textit{comparison array} is shown, and subjects are prompted to respond within 2000 ms if the targets in the comparison array match or mismatch the standard array.  The total duration of the experiment,  which consists of 359 trials, was 5-10 minutes for each subject.  EEG data was downsampled to 256 Hz, channels with corrupted signals removed, and a band pass filter from 1 -- 40 Hz was applied. Independent components analysis was then used to remove physiological artifacts such as eyeblinks and other muscle movement \citealp{delorme2004eeglab, makeig2004mining}.  Lastly, all subjects were manually inspected to remove any remaining artifacts.

\newpage
\begin{table}[H]
\centering
\begin{tabular}{cccccc}
\toprule
$\sigma_A^2$ & AdaMant & Mantel & RV & dCov & aSPC\\
\midrule
 0 &     0.059 &  0.021  & 0.045 &  0.043  & 0.052\\
 0.01 &  0.251 &  0.070 & 0.244 &  0.253 &  0.077\\
 0.02 & 0.759 & 0.115  & 0.756  & 0.752  & 0.177\\
 0.03 & 0.924 & 0.191 & 0.935  & 0.919  & 0.378\\
 0.04 &  0.981 &  0.313 & 0.980  & 0.981 &  0.462\\
 \bottomrule
\end{tabular}
\caption[Comparison of association test power.]{Power results for association testing with AdaMant compared to other testing methods for simulated data generated from the multivariate random effects model with $n = 200$, $p = 40$, $q = 20$.}
\label{tab:sim_mv}
\end{table}

\FloatBarrier

\FloatBarrier

\FloatBarrier


\FloatBarrier

\FloatBarrier

\newpage
\bibliographystyle{unsrtnat}  
\bibliography{wileyNJD-AMA}%




\end{document}